\documentstyle[amstex,12pt]{article}                 
\newlength{\dinwidth}
\newlength{\dinmargin}
\newtheorem{Definition}{Definition}[section]
\newtheorem{Theorem}[Definition]{Theorem}
\newtheorem{Proposition}[Definition]{Proposition}
\newtheorem{Lemma}[Definition]{Lemma}

\newcommand{\su}{{\underline{\sigma}}}

\newcommand{\Bu}{\underline{B}}

\newcommand{\bBox}{{\mbox{\large $\Box$}}}

\newcommand{\CC}{{\mbox{{\small $ \Bbb C$}}}}
\newcommand{\NN}{{\mbox{{\small $ \Bbb N$}}}}
\newcommand{\RR}{{\mbox{{\small $ \Bbb R \,$}}}}

\newcommand{\la}{{\lambda}}                     
\newcommand{\Oo}{{\cal O}}                      
\newcommand{\Op}{{\cal O}'}                     
\newcommand{\Oi}[1]{{\cal O}_{#1}}              
\newcommand{\AO}{{\frak A} ({\cal O})}           
\newcommand{\AOi}[1]{{\frak A} ({\cal O}_{#1})}
\newcommand{\Al}{{\frak A }}                     

\newcommand{\Alu}{\underline{{\frak A}}}         
\newcommand{\AOu}{\underline{{\frak A}}({\cal O})}

\newcommand{\Au}{\underline{A}}                  
\newcommand{\Aul}{{\underline{A}}_{\lambda}}     
\newcommand{\ax}{{\alpha}_{x}}                   
\newcommand{\Ux}{U(x)}                           
\newcommand{\axu}{{\underline{\alpha}}_x}        
\newcommand{\atu}{{\underline{\alpha}}_t}        
\newcommand{\au}{{\underline{\alpha}}}           

\newcommand{\SL}{{\mbox{SL} ( \omega )}}

\newcommand{\TbO}{{\Theta_{\beta, \Oo}}}
\newcommand{\TlblO}{{\Theta_{\la \beta, \la \Oo}}}
\newcommand{\NbO}{{N_{\beta,\Oo} ( \varepsilon )}}
\newcommand{\NlblO}{{N_{\la \beta, \la \Oo} ( \varepsilon )}}
\newcommand{\TNbO}{{\Theta^{(0)}_{\beta, \Oo}}}
\newcommand{\NNbO}{{N^{(0)}_{\beta,\Oo} ( \varepsilon )}}
\newcommand{\Nus}{{\underline{N}_{\beta^{-1} \Oo} ( \varepsilon ) }}
\newcommand{\Nos}{{\overline{N}_{\beta^{-1} \Oo} ( \varepsilon ) }}

\newcommand{\Hh}{{\cal H}}                       
\newcommand{\Kk}{{\cal K}}                       
\newcommand{\BH}{{\cal B}({\cal H})}             

\newcommand{\ou}{{\underline{\omega}}}           
\newcommand{\oul}{{\underline{\omega}}_\la}      
\begin{document}

\title{ Phase Space Properties of Local Observables \\
and Structure of Scaling Limits}
\author{Detlev Buchholz \\
II. Institut f\"ur Theoretische Physik,
Universit\"at Hamburg \\
D-22761 Hamburg, Germany}
\date{DESY 95-095}

\maketitle
%
%
%
\begin{abstract}
\noindent For any given algebra of local observables in
relativistic quantum field theory there exists an
associated scaling algebra which permits one
to introduce renormalization group transformations
and to construct the scaling
(short distance) limit of the theory. On the basis of this result it is
discussed how the phase space properties of a theory determine the
structure of its scaling limit.
Bounds on the number of local degrees of freedom
appearing in the scaling limit are given which allow one to
distinguish between
theories with classical and quantum scaling limits.
The results can also be used to establish
physically significant algebraic properties of
the scaling limit theories, such as the split property.
\end{abstract}
\section{Introduction}

\noindent The general analysis of the structure of
local observables at small spatio--temporal
scales is in several respects an interesting issue.
It is of relevance in the classification of
the possible ultraviolet properties of
local quantum field theories
and
a prerequisite for the proper
description of the particle--like structures
appearing at small scales, such as quarks and
gluons. Moreover,
the short distance analysis is
crucial to the understanding of
symmetries, such as colour, which come to light
only at small scales, and it may be a key to the
reconstruction of the local gauge groups
from the (gauge invariant) observables.

A general framework for the systematic analysis of
these problems has recently been proposed in \cite{BuVe}.
It is based on ideas of
renormalization group theory (cf.\ \cite{Zi} and references
quoted there) which are incorporated into the setting
of local quantum physics by the novel concept
of scaling algebra. Within this framework one can construct
in a model independent way
the scaling (short distance) limit of any given theory
and analyze its structure.

It is the aim of
the present article to review this approach and
to study the relation between the structure of the scaling
limit and the phase space properties of the underlying
theory.
This analysis is carried out in the algebraic
Haag--Kastler framework of
local quantum theory \cite{Ha}.
Since this setting may be less well--known,
we briefly describe here
in conventional field theoretic terms the basic
ideas and main results of the present investigation.

Given any local quantum field theory, we consider the corresponding
observable Wightman
fields, currents etc.\ which act as operator--valued
distributions on the physical Hilbert space. Since we
are interested in the short distance properties of the
theory, we have to study the effect of a change of the
spatio--temporal scale on the observables, while keeping the
scales $c$ of velocity and $\hbar$ of action fixed. In the field
theoretic setting such a change of scale induces transformations
of the underlying observable fields. If $\phi (x)$ is such a
(hermitean) field which,
at the original scale, is localized at the space--time point $x$,
then the corresponding field at other scales is obtained
by setting
\begin{equation}
\phi_\lambda (x) \doteq N_\lambda \, \phi (\lambda x),
\end{equation}
where $\lambda > 0$ is a scaling factor.
We call $\phi_\lambda (x)$ the field at scale $\lambda$.
Whereas the action of the
scaling transformations on the argument of the field needs no
explanation, its effect on the scale of field strength, given
by the positive factor $N_\lambda$, is more subtle. The
familiar idea is to adjust this factor in such a
way that the expectation values of the fields
at scale $\lambda$ in some given reference
state are of equal order of magnitude for all $\lambda > 0$.  The
precise way in which this idea is implemented is a matter of
convention. One may integrate for example $\phi_\lambda (x)$
with a suitable (real) test function $f(x)$,
\begin{equation}
\phi_\lambda (f) \doteq \int \!  d^{\, 4} x f(x) \, \phi_\lambda (x),
\end{equation}
and demand that
\begin{equation}
    ( \Omega, \, \phi_\lambda (f)  \phi_\lambda (f) \, \Omega )
    = \mbox{const} \quad \mbox{for} \quad \lambda > 0,
\end{equation}
where $\Omega$ is the vacuum vector. Yet one could impose
such a constraint just as well on higher
moments of $\phi_\lambda (f)$.
By conditions of this type one
can determine the factor $N_\lambda$ and thereby adjust
the scale of field strength. We denote the
factor $N_\lambda$ which has been fixed
by such a renormalization condition
by $Z_\lambda$.

For the analysis of the theory at small scales one
has to consider the $n$--point  correlation functions
of the fields at
scale $\lambda$ and to
proceed to the scaling limit $\lambda \searrow 0$.
There appears, however,  a problem. As is well known,
the product of quantum fields at neighbouring space--time points is
quite singular and consequently the renormalization factors
$Z_\lambda$ tend to $0$ in this limit.
One needs rather precise information on the way how
$Z_\lambda$ approaches $0$ in order to be able to control
the scaling limit of the fields. It is apparent that if one
chooses in relation (1.1) a factor $N_\lambda$
such that the quotient $N_\lambda / Z_\lambda$ approaches
$0$ or $\infty$ in the scaling limit one ends up with a
senseless result.

In the conventional setting of quantum field theory this problem
can be solved under favourable circumstances (asymptotically free
theories), since renormalization group equations and
perturbative methods provide reliable information on the
asymptotic behaviour of $Z_\lambda$. This approach
works also in some renormalizable theories where
the underlying renormalization group equations have a
non--vanishing but
small ultraviolet fixed point.
Yet in the case of theories without a (small)
ultraviolet fixed point the method does not lead to reliable
results, nor can a rigorous treatment of non--renormalizable theories
be even addressed.
In view of this fact a model--independent
approach to the short distance analysis
of local quantum field theories
would seem to be impossible.

A way out of this problem which is quite simple
has been proposed in \cite{BuVe}. Within the field--theoretical
setting it can be described
as follows. In a first step one proceeds from the unbounded field
operators $\phi_\lambda (f)$ to
corresponding bounded operators, such as
the unitary operators $\exp{( i \phi_\lambda (f))}$.
This has the effect that, irrespective of the choice of the
factor $N_\lambda$ in the
definition of $\phi_\lambda (x)$, there do not
appear any divergence problems: the resulting operators
are bounded in norm, uniformly in $\lambda$.

The second crucial step is to restrict the
four--momentum of these bounded operators
in accord with the uncertainty principle.
Roughly speaking, one considers only
operators which can transfer
to physical states at scale $\lambda$
momentum proportional to $\lambda^{-1}$,
hence they occupy for all $\lambda > 0$
the same phase space volume. The desired operators
are obtained by suitable space--time averages,
\begin{equation}
    A_\lambda \doteq \int \! d^{\, 4} y \, g(y) \, \exp{( i \phi_\lambda
    ( f_y ))},  \quad \lambda > 0,
\end{equation}
where $g(y)$ is any test function and $f_y (x) \doteq f(x-y)$.
It turns out that this restriction
on the momentum transfer has the following effect:
if one chooses $N_\lambda$ such that
$N_\lambda / Z_\lambda$ tends to $\infty$ in the
limit of small $\lambda$, then all correlation functions
involving the corresponding sequence of operators
$A_\lambda$ converge to $0$. Similarly, if
$N_\lambda / Z_\lambda$ tends to $0$, then $A_\lambda$
converges (in the sense of correlation functions)
to $\mbox{const} \cdot  1$. Thus in either
case the operators $A_\lambda$ tend in the
scaling limit to multiples of the identity. Only in the special
case where the asymptotic behaviour of
$N_\lambda$ coincides with that of
$Z_\lambda$ does it happen that the correlation functions
retain a non--trivial operator content in the
scaling limit.

In view of this fact one does not need to know the behaviour
of $Z_\lambda$ and may admit in the above construction
all possible factors $N_\lambda$. The theory takes care by itself
of those choices which are unreasonable and lets them disappear
in the scaling limit. It is only if  $N_\lambda$ has the right
asymptotic behaviour that the sequences $A_\lambda$ give rise to
non--trivial operator limits. One may view this
method as an implicit way of introducing renormalization group
transformations. It allows one to study the short distance properties
of local quantum field theories in a model independent manner.

It is convenient in this analysis to regard the operators
$A_\lambda$ obtained by the above procedure as functions of the
scaling parameter $\lambda$. These
functions form in an obvious way an algebra, the scaling algebra.
For the construction of
the scaling limit of the theory one considers the
expectation values of sums and products of
the operator functions in
the limit $\lambda \searrow 0$,
\begin{equation}
 \lim_{\lambda \searrow 0} \, \,
( \Omega, \, \sum A^{}_\lambda  A^{'}_\lambda \cdots
  A^{''}_\lambda \, \Omega ) =
( \Omega_0, \, \sum A^{}_0  A^{'}_0 \cdots A^{''}_0 \, \Omega_0 ).
\end{equation}
Actually, these limits may only exist for suitable subsequences
of the scaling parameter $\lambda$. We disregard this problem
for the moment but return to it in the main text.
What is of interest here is the fact that the limits
of the correlation functions
determine, by an application of the reconstruction theorem
to the scaling algebra, a local, covariant
theory with unique vacuum vector $\Omega_0$. This explains the
notation on the right hand side of equation (1.5). By this
universal method one can construct the scaling limit of any
given theory.

On the basis of this result the possible structure of
scaling limit theories has been classified in \cite{BuVe}.
There are two extreme cases. The first
possibility is that the scaling limit theory is ``classical''.
This happens if the correlations between all observables
disappear at small scales and the correlation functions
in (1.5) factorize in the limit. Examples may be certain
non--renormalizable theories, where the leading short distance
singularities of the fields are not governed by the two
point functions (cf. also the remarks in the conclusions).
In the second, more familiar case the quantum
correlations persist in the scaling limit. One then ends up with
a full--fledged quantum field theory.

We want to clarify in
this article the relation between the nature of the
scaling limit and the phase space properties of the underlying
theory and establish criteria
for deciding which of the
above cases is realized.
Although phase space is a poorly defined concept in quantum
field theory, one can introduce a measure of its size
by appealing to a semiclassical picture and
counting the
number of states of limited energy which are localized in a fixed
spacetime region.
To illustrate this idea, let us
consider for given spacetime region $\Oo$ and any $\lambda > 0$ the
vectors
\begin{equation}
 \int \! d^{\, 4} y \, g ( y ) \,
 \exp{(i \phi_\lambda ( f_y ))} \, \Omega,  \quad
 \mbox{supp} \, f \subset {\cal O}.
\end{equation}
Here $g$ is a fixed test function whose
Fourier transform has support in a given region
$\widetilde{{\cal O}}$
of momentum space.
These vectors describe local
excitations of the vacuum vector $\Omega$ in the region
$\lambda {\cal O}$ whose energy momentum content is confined
to $\lambda^{-1} \widetilde{{\cal O}}$. In the following
we denote the set of all these vectors by ${\cal S}_\lambda$.

It has been pointed out by Haag and Swieca \cite{HaSw} that,
disregarding vectors of small norm, the subsets
${\cal S}_\lambda$ of the physical Hilbert space
should be finite dimensional
in physically reasonable theories.
A convenient measure of the size of these sets is
provided by the notion of epsilon content. This
is the number $N_\lambda ( \varepsilon )$ of vectors in
${\cal S}_\lambda$ whose mutual distance is larger than a given
$\varepsilon > 0$. Thus
the epsilon content $N_\lambda ( \varepsilon )$ provides
information on the number of states which are
affiliated with the regions
$\lambda {\cal O}$, $\lambda^{-1} \widetilde{{\cal O}}$
of configuration and momentum space. We mention as an aside
that the dependence of these numbers
on $\varepsilon$ and the given
regions is intimately related to thermal
properties of the underlying theory \cite{BuWi,BuJu}.

As we shall see, the structure of the scaling limit of a theory
depends crucially on the size of the epsilon contents $N_\lambda (
\varepsilon )$ in the limit $\lambda \searrow 0$.
Disregarding
fine points, there are the following possibilities: if the limit
$N_0 ( \varepsilon )$ of the epsilon contents behaves for small
$\varepsilon$ like $\varepsilon^{- p} $ for some $p > 0$, then the
scaling limit is classical. Otherwise it is a quantum field
theory. Moreover, if $N_0 ( \varepsilon )$ behaves like
$\exp{(\varepsilon^{-q})}$ for some
sufficiently small $q > 0$, then the scaling limit
theory satisfies a nuclearity condition, proposed in \cite{BuWi},
which is a sharpened version of the Haag--Swieca criterion.
This result can be used to establish a strong form of causal
independence (split property \cite{DoLo}) in the scaling limit
theories. On the other hand,
if the epsilon contents $N_\lambda
( \varepsilon )$ diverge in the limit of small $\lambda$, then the
scaling limit theory no longer complies with the condition of
Haag and Swieca.

Further results of a similar nature which
reveal an intimate connection between
the number of degrees of freedom affiliated with
specific regions of phase space and the
short distance properties of a theory will be given in the main text.
One may hope that
these results provide the basis for
an extensive analysis and classification
of the possible ultraviolet properties of local
observables.

We conclude this introduction with a brief summary.
In the subsequent section we recall the basic notions used
in the algebraic approach to local quantum physics
and compile some
results which enter
in our discussion of phase space properties.
Section 3 contains a review of the construction of scaling
algebras and scaling limits.
The heart of the paper is Section 4, where the
analysis of the relation between phase space and short
distance properties is given.
The paper closes with a brief discussion of examples and an
outlook on further
developments of the theory.
\section{Phase space properties of local observables}
\setcounter{equation}{0}

\noindent As mentioned in the Introduction, we make use in this
investigation of the algebraic framework of local quantum
physics \cite{Ha}. This allows us to discuss the short distance
properties of local observables in full
generality, including also gauge theories, where one considers
besides point like fields other observables, such as
Wilson loops and string operators.
In order to establish our terminology we list in
the first part of this section the basic assumptions
which ought to be satisfied by the observables in any physically
reasonable theory. In the second part
we recall some mathematical concepts and results
which are of relevance in our
discussion of phase space properties.

\vspace{0.5em}
1.  {\em (Locality)\/}
The observables of the underlying theory generate a {\em local
net\/} over Minkowski space $\RR^4$, that is an
inclusion preserving map $\Oo \rightarrow \AO$
from the set of open, bounded regions $\Oo \subset \RR^4$ to
unital C$^*$-algebras $\AO$ on the physical Hilbert
space $\Hh$. Thus each $\AO$ is a norm closed subalgebra of the
algebra $\BH$ of all bounded operators on $\Hh$ which is stable
under taking adjoints and contains the unit operator, and there holds
\begin{equation} \Al (\Oi{1}) \subset \Al (\Oi{2}) \quad \mbox{if}
    \quad \Oi{1}
\subset \Oi{2}. \end{equation}
The net complies with the principle of locality (Einstein
causality) according to which observables in spacelike separated
regions commute,
\begin{equation} \AOi{1} \subset \AOi{2}' \quad \mbox{if} \quad  \Oi{1}
    \subset
\Oi{2}'.\end{equation}
Here
$\Op$
denotes the spacelike complement of
$\Oo$
and
$\AO '$
the set of operators in
$\BH$
which commute with all operators in
$\AO$.
One may think of $\AO$ as the algebra generated by
all observables
which can be measured in the spacetime region $\Oo$.
The global algebra $\Al$ of observables
is generated by all local algebras $\AO$
(as inductive limit in the norm topology).
We recall that for the interpretation of a theory
it is not really necessary to assign a specific physical
meaning to individual operators. All what matters
is the information about the localization properties of
the operators, which is encoded in the net structure \cite{Ha}.

\vspace{0.5em}
2. {\em (Covariance)\/} On the Hilbert space
$\Hh$ there exists a continuous unitary
representation $U$ of the space--time translations $\RR^4$ which
induces automorphisms of the
given net of observables.\footnote{ We make no assumptions with
regard to Lorentz transformations. Thus the present framework is
slightly more general than the one used in \cite{BuVe}.}
Thus for each $x \in \RR^4$ there is
an $\ax \in \mbox{Aut} \Al$ given by
\begin{equation} \ax ( A ) \doteq \Ux A \Ux^{-1}, \quad A \in \Al,
\end{equation}
and, in an obvious notation,
\begin{equation} \ax ( \AO ) = \Al ( \Oo + x ) \end{equation}
for any region
$\Oo$.
In addition to this fundamental postulate we assume that
the operator valued functions
\begin{equation} x \rightarrow \ax (A), \; A \in \Al \end{equation}
are continuous in the norm topology. This condition, which is
crucial in the present investigation, is always satisfied by a
sufficiently rich set of local observables and does not impose any
essential restrictions of generality.

\vspace{0.5em}
3. {\em (Spectrum condition)\/} The joint spectrum of the
energy--momentum operators, i.e., the generators
of the unitary representation $U$ of the translations, is
contained in the closed forward lightcone
$\overline{V}_+ = \{ p \in \RR^4 : p_0 \geq | \mbox{\boldmath
$p$ } | \} $.
Moreover, there is an (up to a phase unique) unit
vector $\Omega \in \Hh$, representing the vacuum, which is invariant
under the action of the representation $U$,
\begin{equation} \Ux \Omega = \Omega, \quad x \in \RR^4. \end{equation}

Let us turn now to the
description of the phase space properties of the
observables. There we have to rely on concepts
from the theory of compact linear maps between
Banach spaces. We recall here some basic definitions and
useful results.

Let ${\cal E}$ be any
Banach space with norm $\| \cdot \|_{\cal E}$. The
unit ball of ${\cal E}$ is denoted by
${\cal E}_{\mbox{{\scriptsize I}}}$ and the space of
continuous linear functionals on ${\cal E}$ by ${\cal E}^{\ast}$. Given
another Banach space ${\cal F}$, we denote the space of continuous
linear maps $L$ from ${\cal E}$ to
${\cal F}$ by ${\cal L} ({\cal E,\cal F})$.
The latter space is again a Banach space with norm given by
\begin{equation}
\| L \| \doteq \sup \{ \| L (E)\|_{\cal F}:E
\in {\cal E}_{\mbox{{\scriptsize I}}} \}.
\end{equation}
A map $L \in {\cal L} ({\cal E, \cal F})$ is said to be compact if the
image of
${\cal E}_{\mbox{{\scriptsize I}}}$
under the action of $L$ has compact closure
in ${\cal F}$. A convenient measure which provides more detailed
information on the size of the range of compact maps is the notion
of epsilon content.
\vspace{0.5em}

\noindent{\em Definition\/}: Let $L\in{\cal L}({\cal E, \cal F})$
and let, for given $\varepsilon > 0$,
$N_L(\varepsilon)$ be
the maximal number of elements
$E_i\in{\cal E}_{\mbox{{\scriptsize I}}}, \, i=1,\,\dots
N_L(\varepsilon)$, such that
$\| L (E_i-E_j)\|>\varepsilon$ for
$i\not= j$. The number $N_L(\varepsilon)$ is called
{\em epsilon content\/} of $L$. It
is finite for all $\varepsilon > 0$ if
and only if $L$ is compact \cite{Pi}.
\vspace{0.5em}

It is apparent that the epsilon content of $L$ increases if
$\varepsilon$
decreases, and it tends to infinity if $\varepsilon$ approaches 0
(unless $L$ is the zero map).
If $L$ is of finite rank, then the
epsilon content $N_L(\varepsilon)$ behaves for small $\varepsilon$
like $\varepsilon^{-p}$ for some
positive number $p$. The converse statement
is also true \cite{Pi}.

We are now in a position to formulate the condition proposed by
Haag and Swieca \cite{HaSw} to characterize physically significant
theories with decent phase space properties.
We state this condition here in a slightly modified
but mathematically equivalent form.

\vspace{0.5em}
4. {\em (Compactness)\/} For given $\beta > 0$
and spacetime region ${\cal O}$,
let $\Theta_{\beta, {\cal O}}$
be the map from $\AO$ into ${\cal H}$,
defined by
\begin{equation}
\Theta_{\beta, {\cal O}} (A) \doteq e^{- \beta H}
A \Omega , \quad A \in \AO ,
\end{equation}
where $H$ is the (positive) generator of the time translations.
The maps $\Theta_{\beta, {\cal O}}$
are compact for any $\beta > 0$ and any bounded ${\cal O}$.
\vspace{0.5em}

For physical motivations of this condition, cf. \cite{HaSw} and
\cite{BuWi}.
The compactness condition has rigorously been established
in several models. Examples are massive \cite{HaSw}
and massless \cite{BuJa} free field theories and certain
interacting theories in two spacetime dimensions, such as
the $P ( \phi )_2$ models, the Yukawa theory $Y_2$ and
theories exhibiting solitons, cf.\ \cite[Sec.\ 4]{BuDaLo} and
references quoted there. In fact, any theory which has the
so--called split property also satisfies the Haag--Swieca
compactness condition \cite{BuDaLo}.
On the other hand one knows that the compactness condition is
violated in theories with an unreasonably
large number of local degrees of freedom,
such as generalized free fields
with continuous mass spectrum
\cite{HaSw}. Thus
the maps $\Theta_{\beta, {\cal O}}$ are
a convenient tool to characterize the
phase space properties of a theory.
For a discussion of related concepts and their comparison
with the compactness condition,
cf.\ \cite{BuDaLo} and \cite{BuPo}.

An important class of compact maps which enter in quantitative
versions of the compactness condition, proposed in
\cite{BuWi} and \cite{BuDaLo}, are the so--called nuclear
maps. They are defined as follows.

\vspace{0.5em}
\noindent{\it Definition\/}: Let $L\in{\cal} L({\cal E}, {\cal F})$ be a
map such that for suitable sequences $e_n \in {\cal E}^{\ast}$
and $F_n\in{\cal F}$, $n\in\NN$,
there holds (in the sense of strong
convergence in ${\cal F}$)
$$
L(E)=\sum_n e_n(E) F_n, \quad E \in {\cal E}.
$$
If there holds in addition
$\sum_n
\| e_n \|_{{\cal E}^{\ast}}^p \| F_n \|_{\cal F}^p <\infty$ for
some $p>0$ the map $L$ is said to be $p$--{\it nuclear\/}. The
$p$--nuclear maps form a vector space
which is equipped with the (quasi) norm \cite{Pi}
$$
\| L \|_p = \inf \big( \sum_n
\| e_n \|_{{\cal E}^{\ast}}^p \| F_n \|_{\cal F}^p
\bigr)^{1/p},
$$
where the infimum is to be taken with respect to all possible
decompositions of $L$. A map which is $p$--nuclear for all
$p > 0$ is said to be of {\em type s\/}.

It has been argued in \cite{BuDaLo} that, disregarding theories
with a maximal (Hagedorn) temperature, the maps $\TbO$ in the
compactness condition ought to be of type {\em s\/}, i.e.,
\begin{equation}
|| \TbO ||_p < \infty \quad \mbox{for} \quad p > 0.
\end{equation} In the present
investigation we will extract from the dependence
of the $p$--norms on $\beta$ and $\Oo$ information on the nature
of the scaling limit of the theory. In the argument we make use
of the fact that the epsilon content and the $p$--norms of nuclear
maps are closely related. We quote in this context the following
useful result.

\begin{Lemma} Let $L \in {\cal L}({\cal E, F})$, where
${\cal F}$ is a Hilbert space. \\
(i) If $L$ is $p$--nuclear for some $0 < p < 1$, its epsilon content
satisfies for any $q > p / (1 - p)$
$$ N_L (\varepsilon) \leq \exp{(c \, || L ||_p^q \, /
\varepsilon^q_{} )},
\quad \varepsilon > 0, $$
where the constant $c$ depends on $p,q$, but not on $L$. \\
(ii) Conversely, if for some $0 < p \leq 1$ there is a sequence
$\varepsilon_m > 0, m \in \NN$, such that
$$ \sum_m \, ( m^{1/2} \varepsilon_m N_L (\varepsilon_m)^{1/m} )^p
< \infty, $$
the map $L$ is $p$--nuclear and
$$ || L ||_p \leq (2 \pi )^{1/2}
\big( \sum_m ( m^{1/2} \varepsilon_m N_L (\varepsilon_m)^{1/m} )^p
\big)^{1/p}.$$
(iii) If, for some $0 < q < 2/3$, there holds
$$ \sup_{\varepsilon > 0} \, \varepsilon
\big( \ln{ N_L ( \varepsilon ) } \big)^{1/q} < \infty, $$
the map $L$ is $p$--nuclear for $p > 2q / (2 - q)$ and
$$ || L ||_p \leq c \, \sup_{\varepsilon > 0} \, \varepsilon
\big( \ln{ N_L ( \varepsilon ) } \big)^{1/q}, $$
where $c$ depends on $p,q$, but not on $L$.
\end{Lemma}

The arguments for the proof of these statements can be extracted
from Proposition~2.5 and Lemma~2.2 in \cite{BuDa}. We refrain from
giving here the straightforward details.
\setcounter{section}{2}
\section{Scaling algebras and scaling limits}
\setcounter{equation}{0}
\noindent
We review in this section the construction of the scaling
algebra associated with any given local net of observables which
complies with the first three conditions given in Sec.\ 2. As has
been shown in \cite{BuVe}, the scaling algebra provides a convenient
tool for the definition and analysis of the scaling limit of a
theory.

The elements of the scaling algebra are bounded functions of the
scaling parameter
$\la > 0$
with values in the algebra of observables
$\Al$,
\begin{equation}  \Au \, : \, \RR^+ \rightarrow \Al. \end{equation}
We mark these functions in the following by underlining. A simple
example of such a function has been given in the Introduction,
cf.\ relation (1.4). As has been discussed in \cite{BuVe}, the
values
$\Au_\la$
of the functions $\Au$
are, for given
$\la > 0$,
to be interpreted as elements of the theory at scale
$\la$.
One therefore defines for these functions the following algebraic
relations: for any
$\Au, \Bu$
and
$a,b \in \CC$
one puts for
$\la>0$
\begin{eqnarray}
 (a\Au + b \Bu)_{\la} & \doteq & a\Aul + b\Bu_{\la} \nonumber \\
 (\Au \cdot \Bu)_{\la} & \doteq& \Aul \cdot \Bu_{\la} \\
 (\Au^*)_{\la} & \doteq & {\Aul}^{\! *} \,. \nonumber
\end{eqnarray}

In this way the functions
$\Au$
acquire the structure of a *-algebra with unit
given by
$\underline{1}_\la = 1$.
A norm on this algebra (which in fact is a C$^*$--norm) is obtained by
setting
\begin{equation} || \Au || \doteq \sup_{\la > 0} \, || \Au_\la ||.
\end{equation}
The translations
$x \in \RR^4$
induce an action
$\axu $
on the functions
$\Au$,
given by
\begin{equation} ( \axu ( \Au ) )_\la \doteq
{\alpha}_{\la x} ( \Au_\la ). \end{equation}
One considers only functions
$\Au$
which satisfy with respect to this action the continuity condition
\begin{equation} || \axu ( \Au ) - \Au || \rightarrow
0 \quad \mbox{for} \quad x \rightarrow 0. \end{equation}
This crucial requirement amounts to specific restrictions on the
four-momentum of the operators
$\Au_\la, \la > 0$,
which are suggested by the basic ideas of renormalization group
theory
\cite{BuVe}. The effect of this restriction has been described in
the Introduction in the case of a simple example (cf.\ the remarks
after relation (1.4)). In \cite{BuVe} it was assumed that also the
Lorentz transformations act norm--continuously on
$\Au$.
But we do not make such an assumption here and consider the larger
class of functions satisfying only condition (3.5).

The local structure of the underlying net of observables induces
a corresponding local structure on the functions
$\Au$. One defines for any open, bounded spacetime region
$\Oo \subset \RR^4$
the subset
$\AOu$
of continuous (in the sense of condition (3.5)) functions given by
\begin{equation} \AOu \doteq \left\{ \Au \, : \,
\Au_\la \in \Al ( \la \Oo ) \quad \mbox{for} \quad \la > 0 \right\}.
\end{equation}
Since each
$\Al ( \la \Oo)$
is a C$^*$--algebra, it follows that $\AOu$ is a
C$^*$--algebra as well. Moreover, there holds
\begin{equation} \Alu ( \Oo_1 ) \subset \Alu ( \Oo_2 ) \quad
\mbox{if} \quad \Oo_1 \subset \Oo_2, \end{equation}
thus the assignment
$\Oo \rightarrow \AOu$
defines a net of C$^*$--algebras over Minkowski space.
The C$^*$--inductive
limit of this net is denoted by
$\Alu$
and called {\em scaling algebra\/}. It is a straightforward
consequence of conditions (2.2) and (2.4) that the net
$\Oo \rightarrow \AOu$
is local,
\begin{equation} \Alu ( \Oo_1 ) \subset \Alu ( \Oo_2 )'
\quad \mbox{if} \quad \Oo_1 \subset {\Oo_2}',            \end{equation}
(where, by abuse of notation, we have used the symbol
$\Alu ( \Oo )'$
for the relative commutant of
$\AOu$
in
$\Alu$)
and covariant,
\begin{equation} \axu ( \AOu ) \, = \, \Alu ( \Oo + x ). \end{equation}
The local, covariant net\footnote{As is common practice, we denote
the net and its inductive limit by the same symbol.}
$\Alu, {\au}_{\mbox{{\footnotesize ${\Bbb R}^4$}}}$
is called {\em scaling net} of the underlying theory.

In this general formalism one can describe changes of the
spatio--temporal scale by an automorphic action
$\su_{\mbox{{\footnotesize ${\Bbb R}^+$}}}$
of the multiplicative group
$\RR^+$
on the scaling algebra
$\Alu$.
It is given for any
$\mu > 0$
by
\begin{equation} ( \su_\mu ( \Au ) )_\la \doteq \Au_{\la \mu},
\quad \la > 0. \end{equation}
As is easily verified, there hold the relations
\begin{eqnarray}
& \su_\mu \mbox{\footnotesize $\circ $ } \axu   =  \au_{\mu x}
\mbox{\footnotesize $\circ \,$} \su_\mu & \nonumber \\
& \su_\mu ( \AOu )  =  \Alu ( \mu \Oo ), &
\end{eqnarray}
which reveal the geometrical significance of the scaling
transformations
$\su_{\mbox{{\footnotesize ${\Bbb R}^+$}}}$.
These automorphisms may be understood as the algebraic version of
renormalization group transformations \cite{BuVe}.

The scaling net comprises in a comprehensive manner information
about the underlying theory at all spatio--temporal scales.
Moreover, it allows one to {\em compare\/} the properties of the
theory at different scales because of the connection between the
respective observables established by the functions
$\Au$.
This connection is not as rigid as in the conventional approach
to the renormalization group, cf.\ relation (1.1). But it contains
sufficient information for the physical interpretation of
the theory at arbitrarily small spatio--temporal scales.

Within the setting of the scaling algebra the short distance
analysis of physical states is performed as follows: if
$\omega$
is any given physical state on the algebra of observables
$\Al$
(e.g., the vacuum state
$\omega ( \cdot ) = ( \Omega, \cdot \, \Omega)$)
one defines its lift
$\oul$
to the scaling algebra
$\Alu$
at given scale
$\la > 0$
according to
\begin{equation} \oul ( \Au ) \doteq \omega ( \Au_\la ), \quad
\Au \in \Alu. \end{equation}
The functionals
$\oul$
are states on the net
$\Alu, \au_{\mbox{{\footnotesize ${\Bbb R}^4$}}}$,
from which one can recover the properties of the given state
$\omega$
at scale
$\la > 0$,
cf. \cite[Prop.3.4]{BuVe}. Moreover the formalism allows one to
proceed to the scaling limit
$\la \searrow 0$.
To this end one has to regard the family of states
$\oul, \la > 0,$
as a net directed towards
$\la = 0$
and to study its limit behaviour.

As has been discussed in \cite{BuVe}, there appears a minor
technical problem: the net
$\oul, \la > 0$,
does not converge since the scaling algebra comprises the orbits
of local observables arising from an abundance of admissible
renormalization group transformations (cf. the freedom of
choosing
$N_\la$
in the example (1.4)). But, being a bounded set of functionals
on the Banach space
$\Alu$,
the net always contains subnets which converge in the
weak-*-topology according to the Banach-Alaoglu theorem. We recall that
the latter statement means that there exist states
$\ou_0$
on
$\Alu$
such that for any given finite set of elements
$\Au^{(n)} \in \Alu, n = 1, \dots N $,
one can find some sequence
$\la_m, m \in \NN$,
tending to $0$, such that
\begin{equation} \lim_{m \rightarrow \infty} \, \ou_{\la_m}
(\Au^{(n)}) = \ou_0 (\Au^{(n)} ) \quad \mbox{for} \quad n = 1, \dots N.
\end{equation}
The set of all scaling limit states
$\ou_0$
on
$\Alu$
which arises from the given state
$\omega$
on
$\Al$
by this construction, is denoted by
$\mbox{SL} ( \omega )$.
We mention as an aside that
$\mbox{SL} ( \omega )$
does not depend on the choice of
$\omega$
within the class of physically admissible states \cite[Cor.\ 4.2]{BuVe}.

Although the set
$\mbox{SL} ( \omega )$
contains many elements, the apparent ambiguities in the definition
of the scaling limit disappear in general if one takes into account
the proper interpretation of the states
$\ou_0$,
inherited from the states
$\ou_\la$
at finite scales
$\la > 0$.
The procedure is as follows. Given
$\ou_0 \in \mbox{SL} ( \omega )$
one applies the GNS--reconstruction theorem, yielding a
representation
$\pi_0$
of
$\Alu$
on some Hilbert space
$\Hh_0$
and a cyclic vector
$\Omega_0$
such that
\begin{equation} \ou_0 ( \Au ) = ( \Omega_0,  \pi_0 ( \Au ) \,
\Omega_0 ), \quad \Au \in \Alu. \end{equation}
It has been shown in \cite[Lem.\ 4.3]{BuVe} that any
$\ou_0 \in \mbox{SL} ( \omega )$
is a pure vacuum state on
$\Alu$.
Hence there exists on
$\Hh_0$
a continuous unitary representation
$U_0$
of the space-time translations
$\RR^4$
such that for
$\Au \in \Alu$
there holds
\begin{equation} U_0 (x) \pi_0 ( \Au ) {U_0 ( x )}^{-1} =
\pi_0 ( \axu ( \Au ) ), \quad x \in \RR^4. \end{equation}
Moreover, the joint spectrum of the generators of
$U_0$
is contained in the closed forward lightcone
$\overline{V}_+$
and
$\Omega_0$
is the (up to a phase) unique unit vector in
$\Hh_0$
which is invariant under the action of
$U_0$,
\begin{equation} U_0 ( x ) \Omega_0 = \Omega_0. \end{equation}
{}From the latter fact it follows that the algebra
$\pi_0 ( \Alu)$
is irreducible.

For the physical interpretation of
$\ou_0$
one proceeds to the corresponding net
\begin{equation} \Oo \rightarrow \Al_0 ( \Oo ) \doteq
\pi_0 ( \AOu ) \end{equation}
which is local and covariant with respect to the automorphic action
${\alpha}^{(0)}_{\mbox{{\footnotesize ${\Bbb R}^4$}}}$
of the space-time translations, given by
\begin{equation} \ax^{(0)} ( \cdot ) \doteq U_0 (x) \,
\cdot \, U_0 (x)^{-1},
\quad x \in \RR^4. \end{equation}
It has been argued in \cite{BuVe} that the nets
$\Al_0, {\alpha}^{(0)}_{\mbox{{\footnotesize ${\Bbb R}^4$}}}$
obtained in this way for different choices of
$\ou_0 \in \mbox{SL} ( \omega )$
ought to describe the same physics in generic cases, i.e., they
should be isomorphic. One can then regard any one of these nets
as the {\em unique\/} scaling limit of the underlying theory. But
it may also happen that the nets
$\Al_0,  {\alpha}^{(0)}_{\mbox{{\footnotesize ${\Bbb R}^4$}}}$
are non--isomorphic in certain theories
for different choices of
$\ou_0 \in \mbox{SL} ( \omega )$.
This situation will occur if the underlying theory cannot be
described at small scales in terms of a single theory since its
structure varies continually if one approaches
$\la = 0$.
The theory is then said to have a {\em degenerate\/} scaling limit.

In either case the states
$\ou_0 \in \mbox{SL} ( \omega )$
give rise to pure vacuum states
$\omega_0$
on the corresponding nets
$\Al_0, \, {\alpha}^{(0)}_{\mbox{{\footnotesize ${\Bbb R}^4$}}}$,
which are given by
\begin{equation} \omega_0 ( \cdot ) = ( \Omega_0, \, \cdot \, \,
\Omega_0 ). \end{equation}
They describe the properties of the underlying state
$\omega$
on
$\Al$
in the scaling limit. Thus each triple
$\Al_0,  {\alpha}^{(0)}_{\mbox{{\footnotesize ${\Bbb R}^4$}}},
\Omega_0$
complies with the general conditions imposed on a net of observables
in Sec.\ 2, with the possible exception of the compactness
condition.

We are interested here in the general structure of the scaling
limit theories. As has been mentioned in the Introduction, there
exist two clear-cut
alternatives which follow from the fact that the scaling limit states
are pure vacuum states. Given
$\ou_0 \in \mbox{SL} ( \omega )$
one has as a first possibility:  \\[0.5em]
(i)  The net
$\Al_0$ fixed by $\ou_0$ consists only of multiples of the identity.
\\[0.5em]
If this case is at hand for every choice of
$\ou_0 \in \mbox{SL} ( \omega )$
the theory is said to have a {\em classical}  scaling limit.
For in such theories all correlations between observables
disappear at small scales. In theories with a degenerate
scaling limit it may happen, however, that only some of the
states in
$\mbox{SL} ( \omega )$
lead to nets which belong to the
preceding class. The second possibility is: \\[0.5em]
(ii) The net $\Al_0$
associated with
$\ou_0$
is non--trivial in the sense that the algebras corresponding to
(sufficiently large) bounded regions are infinite dimensional
and non--commutative. \\[0.5em]
That the latter case is the only alternative to
the former can be seen in many ways. It follows for example
from the following lemma which we quote for later reference.
Its proof is based on standard arguments.

\begin{Lemma} Let $U$ be a continuous unitary representation of
$\RR^4$
on a Hilbert space
$\Hh$
which satisfies the relativistic spectrum condition, let
$\Omega \in \Hh$
be an (up to a phase unique) unit vector which is invariant under
the action of $U$ and let $A$ be a bounded operator on
$\Hh$
such that
$[ A^* , \, U(x) A U(x)^{-1}] = 0$
for $x$ varying in some open set of
$\RR^4$.
There are the following two alternatives. \\
(i) $A \Omega = a \, \Omega$ for some $a \in \CC$. \\
(ii) For any $r > 0$ the linear span of the vectors
$ U(x) A \Omega, \, |x| < r,$
is infinite dimensional. Then the same holds true for the *-algebra
generated by the operators
$U(x) A U(x)^{-1}, \, |x| < r$,
and this algebra is non--commutative if $r$ is sufficiently large.
\end{Lemma}
\noindent {\em Proof:\/}
It follows from the spectrum condition by an
argument of the Reeh--Schlieder
type that the closure
$\Kk$
of the linear span of vectors
$ U(x) A \Omega, \, |x| < r,$
coincides with that of the vectors
$ U(x) A \Omega, \, x \in \RR^4$.
Thus
$\Kk$
is invariant under the action of $U$. If
$\Kk$
is finite dimensional one can see by an application of the spectral
theorem to the unitary group
$U(\RR^4) | \Kk$
that
\begin{equation}
( A \Omega, \, U(x) A \Omega) = \sum_n \, a_n e^{i p_n x}, \quad
x \in \RR^4. \end{equation}
Here the sum is finite and
$ p_n \in \overline{V}_+$
because of the spectrum condition. By the
commutation properties of $A$ and the invariance of $\Omega$ under
the action of $U$, this function coincides with
$(A^* \Omega, \, U(-x) A^* \Omega)$
for $x$ varying in some open set. The latter function extends, because
of the spectrum condition, to an analytic function on the
backward tube
$\RR^4 -i {V}_+$
and it is bounded there. Since the former function is entire
analytic the edge of the wedge theorem implies that the two
functions coincide on the backward tube. Hence, in view of their
boundedness, there holds
$p_n = 0$,
i.e., the functions are constant. Consequently
$U(x) A \Omega = A \Omega$ for $x \in \RR^4$,
and taking into account the uniqueness of the invariant vector
$\Omega$
one arrives at the conclusion that
$A \Omega = a \, \Omega$
for some
$a \in \CC$. This is case (i).

The alternative is that
$\Kk$
is infinite dimensional. Then the *-algebra generated by the
operators
$U(x) A U(x)^{-1}, \, |x| < r$,
is infinite dimensional as well since the subspace obtained by
applying this algebra to
$\Omega$
contains
$\Kk$.
Moreover, if this algebra were commutative for every choice of
$r > 0$,
the commutator
$[ A^* , \, U(x) A U(x)^{-1}]$
would vanish for all
$x \in \RR^4$
and consequently the functions
$x \rightarrow (A \Omega, \, U(x) A \Omega )$
and
$x \rightarrow (A^* \Omega, \, U(-x) A^* \Omega)$
would coincide. Since the Fourier transforms of these functions
have support in
$\overline{V}_+$ and $- \overline{V}_+$,
respectively, they would have to be constant and consequently
$A \Omega = a \, \Omega$
for some
$a \in \CC$.
Thus
$\Kk$
would be one--dimensional, which is a contradiction. $\bBox$

Since any local $A$ operator complies with the assumption in this lemma
and since the scaling limit states are pure and hence
(in their superselection sector) unique vacuum states, we see that there
are only the two general possibilities for the nets
$\Al_0$,
mentioned above. If the second case is at hand for {\em every\/}
choice of
$\ou_0 \in \mbox{SL} ( \omega )$
we say the theory has a {\em pure quantum} scaling limit (which may
be degenerate, though).

The preceding classification of the possible structure of the
scaling limit arises as a logical alternative within the general setting
of the theory of local observables. Yet it does not shed any light
on the question as to which case is at hand in a given theory.
This point will be clarified in the subsequent section.
\setcounter{section}{3}
\section{Phase space and scaling limit}
\setcounter{equation}{0}
We turn now to the analysis of the relation between the phase--space
properties of a theory and the nature of its scaling limit. As was
explained in Sec.\ 3, the phase space properties can be described in
terms of the maps
$\TbO$,
defined in relation (2.8), which depend on the choice of a parameter
$\beta > 0$
and a bounded spacetime region
$\Oo \subset \RR^4$.
Throughout this section we assume that these maps are compact and
denote their epsilon contents by
$\NbO, \varepsilon > 0$.

We also consider the analogous maps in the scaling limit theories:
let
$\ou_0 \in \SL$,
let
$\Al_0, \alpha^{(0)}_{{\mbox{\footnotesize ${\Bbb R} $}}^4}$
be the corresponding covariant net and let
$\Omega_0 \in \Hh_0$
be the corresponding vacuum vector. Given
$\beta, \Oo$
we define a map
$\TNbO$
from
$\Al_0 ( \Oo )$
into
$\Hh_0$,
setting
\begin{equation} \TNbO ( A ) \doteq e^{\, - \beta H_0} A \Omega_0, \quad
A \in \Al_0 ( \Oo ), \end{equation}
where
$H_0$
denotes the generator of the time translations in the scaling limit
theory, cf.\ relation (3.15). The epsilon content of this map is
denoted by
$\NNbO, \varepsilon >0$,
provided it is finite.

It is our aim to derive information on the properties of the maps
$\TNbO$ from the structure of the maps $\TbO$
in the underlying theory. In a first preparatory step we
pick a test function
$ f_\beta$ on $\RR$
whose Fourier transform
$\widetilde{f_\beta}$
is equal to
$ (2 \pi)^{- 1/2} \, e^{ - \beta \omega } $
for
$\omega \geq 0$
and arbitrary otherwise. Since the time translations
$\atu$
act norm continuously on the scaling algebra
$\Alu$
it follows that the integrals (in the sense of Bochner)
\begin{equation} \au_{f_\beta} ( \Au ) \doteq
\int dt \, f_\beta (t) \,
\au_t ( \Au ), \quad \Au \in \Alu \end{equation}
are elements of the scaling algebra
$\Alu$.
Each
$\ou_0 \in \SL$
is a weak-*-limit point of the family of states
$\ou_\la, \la > 0$,
which are the lifts of the vacuum state
$ \omega ( \cdot ) = ( \Omega, \cdot \, \Omega ) $
on
$\Al$
to the scaling algebra. Hence, recalling relation (3.13), there exists
for each {\em finite} set of elements
$\Au \in \Alu$
some sequence
$\la_m, m \in \NN$,
tending to $0$, such that
\begin{eqnarray}
\lefteqn{|| e^{ - \beta H_0 } \pi_0 ( \Au ) \Omega_0 ||^2 =
|| \pi_0 ( \au_{f_\beta} ( \Au ) ) \Omega_0 ||^2 }  \nonumber \\
& & = \ou_0 ( \au_{f_\beta} ( \Au )^* \au_{f_\beta} ( \Au ) )
 = \lim_{m} \, \ou_{\la_m} ( \au_{f_\beta} ( \Au )^* \au_{f_\beta}
( \Au ) ) \nonumber \\
& & = \lim_{m} \, ||  \au_{f_\beta} ( \Au )_{\la_m} \Omega||^2 =
\lim_{m} \, || e^{ - \la_m \beta H} \Au_{\la_m} \Omega||^2,
\end{eqnarray}
where in the first and last equality we made use of the spectrum
condition and the specific form of
$f_\beta$.

Relation (4.3) will be the key to the proof of the subsequent
lemmas. In the argument we make also use of the following
well--known fact in the theory of C$^*$--algebras, cf.\ for example
\cite[Ch.\ I.8]{Ta}: if
${\frak B}$
is a unital C$^*$--algebra and
$\pi$
some representation of
${\frak B}$
there holds
\begin{equation} \pi ( {\frak B} )_{\mbox{\scriptsize I}} =
\pi ( {\frak B}_{\mbox{\scriptsize I}} ),
\end{equation}
where the subscript
$\mbox{I}$ denotes the unit ball in the respective algebra.

\begin{Lemma} The epsilon contents of the maps $\TbO$ and $\TNbO$
satisfy
$$ \NNbO \leq \limsup_{\la \searrow 0} \, \NlblO, \quad
\varepsilon > 0. $$ \end{Lemma}
\noindent {\em Proof:\/} Let
$A^{(n)} \in \Al_0 ( \Oo )_{\mbox{{\scriptsize I}}},
n = 1, \dots N$, be such that
$ || \TNbO ( A^{(n')} - A^{(n'')} ) || > \varepsilon $
for
$n' \neq n''$.
As
$\Al_0 (\Oo )_{\mbox{\scriptsize I}} =
\pi_0 ( \AOu )_{\mbox{\scriptsize I}} =
\pi_0 ( \AOu_{\mbox{{\scriptsize I}}} )$,
there exist $N$ elements
$\Au^{(n)} \in \AOu_{\mbox{\scriptsize I}} $,
such that
$A^{(n)} = \pi_0 ( \Au^{(n)} ), n = 1, \dots N$,
and consequently
$$ || e^{ - \beta H_0 } \pi_0 ( \Au^{(n')} - \Au^{(n'')} ) \Omega_0
|| > \varepsilon \quad \mbox{if} \quad n' \neq n''. $$
Since we are only dealing with a finite number of elements of
$\AOu$
we can apply relation (4.3) and find that there is some number
$m_0$
such that for all
$m \geq m_0 $ and $n', n'' = 1, \dots N$ there holds
$$ || e^{ - \la_m \beta H} ( \Au^{(n')}_{\la_m} - \Au^{(n'')}_{\la_m})
\Omega || > \varepsilon \quad \mbox{if} \quad n' \neq n''. $$
But
$ \Au^{(n)}_\la \in \Al ( \la \Oo )_{\mbox{\scriptsize I}}, \la > 0$,
hence we see from the latter inequality that $N$ cannot be larger than
the epsilon content of the maps
$\Theta_{\la_m \beta, \la_m \Oo}, m \geq m_0$.
The statement then follows. $\bBox$

In the next step we establish a lower bound on the epsilon content
of the maps
$\TNbO$
in the scaling limit theories.

\begin{Lemma} Let
$\Oo_0$
be any spacetime region whose closure is contained in the interior
of
$\Oo$
and let
$\beta_0 > \beta$.
The epsilon contents of the maps $\TNbO$ and $\TbO$ satisfy
$$ \liminf_{\la \searrow 0} \, N_{\la \beta_0, \la \Oo_0} \leq
\NNbO, \quad \varepsilon > 0. $$ \end{Lemma}
\noindent {\em Remark:\/} It is an open problem whether this result
persists if one
requires that the elements of the scaling algebra are also
norm--continuous with respect to the action of Lorentz
transformations.

\noindent {\em Proof:\/} Let
$\varepsilon > 0$
and let $N$ be any natural number which is less than or equal to the
limes inferior of the numbers
$N_{\la \beta_0, \la \Oo_0} (\varepsilon )$
for
$\la$
tending to $0$. Accordingly there exist a
$\la_0 > 0$
and, for any given
$0 < \la < \la_0$,
$N$ operators
$A^{(n)}_\la \in \Al ( \la \Oo_0 )_{\mbox{\scriptsize I}}, n=1,\dots N$,
such that
$$ || e^{ - \la \beta H} ( A^{(n')}_{\la} - A^{(n'')}_{\la})
\Omega || > \varepsilon \quad \mbox{if} \quad n' \neq n''. $$
We pick now a suitable non--negative, integrable function $f$
on $\RR^4$ which
has support in a sufficiently small neighbourhood of the origin
$0$
such that
$ \Oo_0 + \mbox{supp} f \subset \Oo$,
in an obvious notation. Moreover,
$\int d^{\, 4} x \, f(x) = 1$,
and the Fourier transform of $f$ has to satisfy the lower bound
\begin{equation}
| \widetilde{f} ( p )| \geq ( 2 \pi )^{-2} \, e^{ - (\beta_0 - \beta)
p_0} \quad \mbox{for} \quad p \in \overline{V}_+. \end{equation}
(Such functions $f$ can be obtained from the elementary function
$\RR \ni u \rightarrow l_r (u)$,
where
$l_r (u) = (1/2r) \ln{(r/|u| )}$ for $|u| < r$
and
$l_r ( u ) = 0$
for
$|u| \geq r$,
setting
$f(x) \doteq \prod_{\nu = 0}^3 \, l_r ( x_\nu )$
and choosing $r$ sufficiently small.) With the help of this function
and the operators
$A^{(n)}_\la$
we can define elements
$\Au^{(n)}, n=1, \dots N$,
of the scaling algebra
$\Alu$
according to
$$ \Au^{(n)}_\la \doteq \int d^{\, 4} x \, f(x) \alpha_{\la x}
( A^{(n)}_\la ) \quad \mbox{for} \quad 0 < \la \leq \la_0, $$
and
$ \Au^{(n)}_\la = 0$
for
$\la > \la_0$.
Since $f$ is absolutely integrable and the operators
$A^{(n)}_\la$
are bounded in norm by $1$ each
$\Au^{(n)}$
satisfies the continuity requirement (3.5). In fact,
$$ || \axu ( \Au^{(n)} ) - \Au^{(n)} || \leq
\int d^{\, 4} x' \, | f(x' - x) - f(x')|. $$
Moreover, because of the localization and normalization properties
of $f$ as well as of the operators
$A^{(n)}_\la$,
there holds
$\Au^{(n)}_\la \in \Al ( \la \Oo )_{\mbox{\scriptsize I}}, \la > 0$,
and consequently
$\Au^{(n)} \in \AOu_{\mbox{\scriptsize I}}, n=1, \dots N$.
We apply once again relation (4.3), giving for
$n', n'' = 1, \dots N, n' \neq n''$,
and some suitable sequence
$\la_m, m \in \NN$,
tending to $0$,
\begin{eqnarray} \lefteqn{
|| e^{- \beta H_0 } \pi_0 ( \Au^{(n')} - \Au^{(n'')} ) \Omega_0 || = }
\nonumber \\
& & = \lim_{m} \, || e^{- \la_m \beta H} ( \Au^{(n')}_{\la_m} -
\Au^{(n'')}_{\la_m} ) \Omega ||  \nonumber \\
& & = \lim_{m} || (2 \pi)^2 \, \widetilde{f} ( \la_m P )
e^{- \la_m \beta H}
( A^{(n')}_{\la_m} - A^{(n'')}_{\la_m} ) \Omega || \nonumber \\
& & \geq
\liminf_{\la \searrow 0} || e^{ - \la \beta_0 H} ( A^{(n')}_\la -
A^{(n'')}_\la ) \Omega || > \varepsilon. \nonumber \end{eqnarray}
Here $P$ is the four--momentum operator and we made use of the lower
bound (4.5) imposed on
$\widetilde{f}$
on the forward lightcone
$\overline{V}_+ $,
which contains the spectrum of $P$. {}From the above estimate and the
fact that
$\pi( \Au^{(n)}) \in \Al_0 ( \Oo )_{\mbox{\scriptsize I}}, n=1,\dots N$,
we see that
$N \leq \NNbO$.
This completes the proof of the statement. $\bBox$

Making use of these results we will establish now conditions in
terms of the epsilon contents
$\NbO, \varepsilon > 0$,
which provide information on the nature of the scaling limit.
In order to simplify the notation we set for $\varepsilon > 0$
\begin{eqnarray} & \Nus \doteq \liminf_{\la \searrow 0} \, \NlblO &
\nonumber \\ & \Nos \doteq \limsup_{\la \searrow 0} \NlblO, &
\end{eqnarray}
provided the respective limits exist. In these definitions we made
use of the obvious fact that the limits depend on
$\beta, \Oo$
in the combination
$\beta^{-1} \Oo$.
The information contained in the preceding two lemmas can thus be
summarized in the inequalities
\begin{equation} \underline{N}_{\beta_0^{-1} \Oo_0} \leq \NNbO \leq
\Nos, \quad \varepsilon > 0, \end{equation}
which will be used in the proof of the following proposition. For
the sake of simplicity we do not aim here at an optimal result. (Cf.
the proof of Proposition 4.5 for certain refinements.)

\begin{Proposition} The following conditions are necessary, respectively
sufficient, for the underlying theory to have a scaling limit of the
given type. \\
(i) Pure quantum scaling limit: it is necessary that there is some
bounded region
$\Oo$
such that for any
$p > 0$
there holds $\varepsilon^p \, \Nos \rightarrow \infty$ as
$\varepsilon \searrow 0$. It is sufficient that
$\varepsilon^2 \,\Nus \rightarrow \infty$ as $\varepsilon \searrow 0$.\\
(ii)  Scaling limit of Haag--Swieca type (all triples
$\Al_0, {\alpha}^{(0)}_{{\mbox{\footnotesize ${\Bbb R}$}}^4}, \Omega_0 $
arising from states in $\SL$ satisfy the Haag-Swieca compactness
condition): it is necessary that $\Nus < \infty, \varepsilon > 0$,
and sufficient that $\Nos < \infty, \varepsilon > 0$,
for all bounded regions $\Oo$. \\
(iii) Classical scaling limit: it is necessary that, for each
bounded spacetime region $\Oo$,
$\varepsilon^2 \, \Nus < \mbox{{\rm const}}$ as $\varepsilon
\searrow 0$. It is sufficient that there is some
$p > 0$ such that $\varepsilon^p \, \Nos < \mbox{{\rm const}}$
as $\varepsilon \searrow 0$. \end{Proposition}
\noindent {\em Proof:\/} In the proof of these statements we make use of
the fact that if $L$ is a (non--zero) compact linear map between Banach
spaces with epsilon content
$N_L (\varepsilon ), \varepsilon > 0$, then
$\varepsilon^p \, N_L ( \varepsilon ) < \mbox{{const}}$
as
$\varepsilon \searrow 0$
for some
$p > 0$
if and only if $L$ has finite rank. In particular,
$\varepsilon^2 \, N_L ( \varepsilon ) < \mbox{{const}}$
as
$\varepsilon \searrow 0$
if and only if $L$ has rank $1$ as a complex linear map. The proof of
the if--part of these statements is straightforward, the
only--if--part is an easy consequence
of the results in \cite[Sec.\ 9.6]{Pi}.

(i) If the theory has a pure quantum scaling limit there
exists
for each triple
$\Al_0, {\alpha}^{(0)}_{{\mbox{\footnotesize ${\Bbb R}$}}^4}, \Omega_0 $
a bounded spacetime region
$ \Oo $
such that the space of vectors
$ \Al_0 ( \Oo ) \Omega$
is infinite dimensional, cf.\ Lemma 3.1. Since
$e^{- \beta H_0}$
is invertible, the space
$e^{- \beta H_0} \Al_0 ( \Oo ) \Omega$
is then infinite dimensional as well, thus the map
$\TNbO$
has infinite rank. Consequently
$\varepsilon^p \, \NNbO \rightarrow \infty$ as
$\varepsilon \searrow 0$ for any finite
$p > 0$.
Because of relation (4.7) this proves the necessity of the
given condition. On the other hand, if for some region
$\Oo_0$
there holds
$\varepsilon^2 \, \underline{N}_{\beta_{0}^{-1} \Oo_0} ( \varepsilon )
\rightarrow \infty$
as $\varepsilon \searrow 0$,
relation (4.7) implies that
$\varepsilon^2 \, \NNbO \rightarrow \infty$ as
$\varepsilon \searrow 0$, provided
$0 < \beta < \beta_0$ and
$\Oo$
contains the closure of
$\Oo_0$ in its interior. Thus the map
$\TNbO$ has rank larger than $1$. {}From Lemma 3.1 and the
covariance of the net
$ \Al_0, {\alpha}^{(0)}_{{\mbox{\footnotesize ${\Bbb R}$}}^4} $
it then follows that the maps
$\Theta^{(0)}_{\beta, \Oo_1}$
have infinite rank for any region
$\Oo_1$
containing the closure of
$\Oo$
in its interior. This proves the sufficiency of the given
condition.

(ii)  The theory has a scaling limit of Haag--Swieca type if
and only if the maps
$\TNbO$ are compact for all
$\beta > 0$ and bounded regions
$\Oo$, i.e., if and only if
$\NNbO < \infty, \varepsilon > 0$.
The statement therefore follows from relation (4.7).

(iii) If the theory has a classical scaling limit all maps
$\TNbO$ are of rank $1$ and consequently
$\varepsilon^2 \, \NNbO < \mbox{{const}}$ as
$\varepsilon \searrow 0$.
The necessity of the given condition then follows from
relation (4.7). On the other hand, if for some
$p > 0, \varepsilon^p \Nos \, < \mbox{{const}}$ as
$\varepsilon \searrow 0$,
there holds also
$\varepsilon^p \, \NNbO < \mbox{{const}}$ as
$\varepsilon \searrow 0$ and consequently the map
$\TNbO$
has finite rank. But in view of Lemma 3.1 this is impossible
for arbitrary bounded regions
$\Oo$
unless all maps
$\TNbO$
are of rank $1$. This proves the sufficiency of the stated
condition. $\bBox$

We mention as an aside that in the interesting case (ii) of this
proposition (scaling limit of Haag--Swieca type) the representation
spaces
$\Hh_0$
of the scaling limit theories
$\Al_0, {\alpha}^{(0)}_{{\mbox{\footnotesize ${\Bbb R}$}}^4}, \Omega_0 $
are separable. This can be seen from the fact that the
countable union of compact sets
\begin{equation} \bigcup_{n \in \mbox{\footnotesize $\Bbb N$}}
n \, e^{- ( \beta /n) H_0} \,
\Al_0 ( \Oo_n )_{\mbox{\scriptsize I}} \, \Omega_0  \end{equation}
is dense in
$\Hh_0$
if the bounded regions
$\Oo_n$
tend to
$\RR^4$
in the limit of large $n$.

Another class of criteria characterizing the nature of the
scaling limit of a theory is obtained by looking at the dependence
of
$\Nus$
and
$\Nos$
on the spacetime region
$\Oo$.
As is apparent from the definition, these quantities depend, for
fixed
$\varepsilon$
and
$\beta$,
monotonically on
$\Oo$.
We make use of this fact in the formulation of the subsequent
result.

\begin{Proposition} The underlying theory has a \\
(i)  pure quantum scaling limit if, for some
$\varepsilon > 0$, $\Nus \rightarrow \infty$ as
$\Oo \nearrow \RR^4$. \\
(ii) classical scaling limit if, for some
$0 < \varepsilon < 2^{1/2}$, $\Nos < \mbox{{const}}$ as
$\Oo \nearrow \RR^4$. \end{Proposition}

\noindent {\em Proof:\/} (i) If $L$ is a complex linear map of rank
$1$ between two Banach spaces and if
$|| L || \leq 1$, its epsilon content satisfies
$ N_L ( \varepsilon ) \leq (1 + \varepsilon^{-1} )^2, \varepsilon > 0.$
This is a straightforward consequence of the fact that the image
of the unit ball under the action of $L$ can be identified with a
circle of radius  $||L||$ in the complex plane.

The defining relation (4.1) and the spectrum
condition imply that
$|| \TNbO || \leq 1$.
Hence if for some triple
$\Al_0, {\alpha}^{(0)}_{{\mbox{\footnotesize ${\Bbb R}$}}^4}, \Omega_0 $
the maps
$\TNbO$
would be of rank $1$ for any
$\beta > 0$
and any bounded region
$\Oo$,
it would follow that
$\NNbO \leq ( 1 + \varepsilon^{-1} )^2$
and consequently
$\lim_{\Oo \nearrow \mbox{\footnotesize $\Bbb R$}^4} \, \Nus < \infty,
\varepsilon > 0$,
because of relation (4.7).
This shows that if the condition in the first
part of the statement is
satisfied, not all of the maps
$\TNbO$
can be of rank $1$. Thus, as was
explained before, the underlying theory has a pure quantum scaling
limit.

(ii) If the scaling limit theory is not classical, there is a
non--trivial scaling limit net
$ \Al_0, {\alpha}^{(0)}_{{\mbox{\footnotesize ${\Bbb R}$}}^4}  $
acting on an infinite dimensional Hilbert space
$\Hh_0$.
Whence, given
$0 < \varepsilon < 2^{1/2}$
and any finite number $N$, there exist a
$\beta_0 > 0$
and $N$ unit vectors
$\Phi_n \in \Hh_0, n = 1, \dots N$,
such that
$$ || e^{ - \beta_0 H_0 } \, ( \Phi_{n'} - \Phi_{n''} ) ||
> \varepsilon \quad \mbox{if} \quad n' \neq n''. $$
This assertion is a simple consequence of the facts that
$e^{- \beta H_0}$
tends to $1$ in the strong operator topology if
$\beta $
tends to $0$ and the norm distance of orthogonal unit
vectors is equal to
$2^{1/2}$. Now since
$\Al_0$
acts irreducibly on
$\Hh_0$
there exist by Kaplansky's density theorem \cite{Sa} some
bounded region
$\Oo_0$
and $N$ operators
$A^{(n)} \in \Al_0 ( \Oo_0 )_{\mbox{\scriptsize I}} $
such that the norm distances
$|| e^{ - \beta_0 H_0 } ( \Phi_n - A^{(n)} \Omega_0 ) ||$
are so small that
$$ || e^{ - \beta_0 H_0 } ( A^{(n')} - A^{(n'')} ) \Omega_0 || >
\varepsilon \quad \mbox{{if}} \quad n' \neq n''.$$
It follows that
$ N^{(0)}_{\beta_0, \Oo_0} ( \varepsilon ) \geq N$
and, by relation (4.7),
$\overline{N}_{\beta_0^{-1} \Oo_0} \geq N$.
Consequently
$\Nos \geq N$
as
$\Oo \nearrow \RR^4$.
Since $N$ was arbitrary,
$\Nos$
diverges
as $\Oo \nearrow \RR^4$
for any choice of
$0 < \varepsilon < 2^{1/2}$.
Thus we conclude that the condition in the second part of the
statement can only be satisfied if the theory has a classical
scaling limit. $\bBox$

Since in physically relevant theories the maps
$\TbO$
are expected to be not only compact but also nuclear
\cite{BuWi, BuDaLo}, the following
results involving the nuclear  $p$--norms
$|| \TbO ||_p$
of these maps are of interest.

\begin{Theorem} Given a theory where the maps
$\TbO$,
defined in (2.8), are $p$--nuclear for some
$0 < p < 1/3$
and
$ \limsup_{\la \searrow 0} || \TlblO ||_p < \infty. $
The theory has a classical scaling limit if and only if there exists
a constant $c$ such that
$$\limsup_{\la \searrow 0} \, || \TlblO ||_{2p} < c$$  uniformly
for all bounded regions
$\Oo$. \end{Theorem}
\noindent {\em Proof:\/} For the proof of the if--part of the statement
we make use of
the first part of Lemma 2.1 which, under the given conditions, implies
that for
$q > 2p / (1 - 2p)$
$$ \Nos \leq \exp{( \mbox{const} / \, \varepsilon^q )}, $$
uniformly in
$\Oo$.
Hence the theory has a classical scaling limit according to the second
part of Proposition 4.4.

For the proof of the only--if--part we have to rely on the following
more refined version of Lemma 4.2: let
$\beta_0, \Oo_0, \varepsilon_0$
be fixed and let
$\la_m, m \in \NN$,
be some sequence, tending to $0$ such that
the sequence of epsilon contents
$N_{\la_m \beta_0, \la_m \Oo_0} ( \varepsilon_0 ), m \in \NN$,
converges or tends to $+ \infty$.
The corresponding subnet of lifted and scaled vacuum
states
$\ou_{\la_m}, m \in \NN$,
still has limit points
$\ou_0 \in \SL$.
For the epsilon contents of the resulting maps
$\TNbO$
one obtains
the estimate (using the same arguments and notation as in the proof
of Lemma 4.2)
$$\lim_{m} \,
N_{\la_m \beta_0, \la_m \Oo_0} ( \varepsilon_0 ) \leq
N^{(0)}_{\beta, \Oo} ( \varepsilon_0 ). $$
Hence if there exist
$\beta_0, \Oo_0, \varepsilon_0$
and a sequence $\la_m, m \in \NN$, as just described,
for which
$$\lim_{m} \,
N_{\la_m \beta_0, \la_m \Oo_0} ( \varepsilon_0 ) >
( 1 + \varepsilon_0^{-1} )^2,$$
the maps
$\TNbO$
cannot be of rank $1$ for all
$\beta, \Oo$.
Assuming that the theory has a classical scaling limit it
follows from this remark and Lemma 3.1 that
\begin{equation} \limsup_{\la \searrow 0} \, \NlblO \leq
( 1 + \varepsilon^{-1} )^2 \end{equation}
for all
$\beta, \Oo$
and
$\varepsilon > 0$.

Now according to the second part of Lemma 2.1 there holds
\begin{equation} \limsup_{\la \searrow 0} \, || \TlblO ||_r
\leq \limsup_{\la \searrow 0} \, ( 2 \pi )^{1/2}
\big( \sum_m ( m^{1/2} \varepsilon_m N_{\la \beta, \la \Oo}
( \varepsilon_m )^{1/m} )^r \big)^{1/r}, \end{equation}
provided the right hand side of this inequality exists for some
$ 0 < r \leq 1 $
and a suitable sequence
$\varepsilon_m, m \in \NN$.
Since all maps
$\TlblO$
are $p$--nuclear for some
$0 < p < 1/3$
this condition is satisfied. In fact, putting
$\varepsilon_m = m^{- 1/q}$,
where
$q > p/(1-p)$,
it follows from the assumptions
and the first part of Lemma 2.1 that
$$ \varepsilon_m N_{\la \beta, \la \Oo}
( \varepsilon_m )^{1/m} \leq c \, m^{- 1/q}, \quad m \in \NN, $$
where the constant $c$ does not depend on
$\la, m$.
Putting
$q=3p/2, r=2p$,
we conclude that the limes superior on the right hand side
of the estimate (4.10) can be pulled under the sum, thereby
leading to a larger upper bound on the left hand side. Hence, taking
into account relation (4.9), we arrive at
$$ \limsup_{\la \searrow 0} \, || \TlblO ||_{2p} \leq
( 2 \pi )^{1/2} \big( \sum_m m^{p - 4/3} ( 1 + m ^{2/3p} )^{4p/m}
\big)^{1/2p}. $$
Since the right hand side of this inequality is finite and does not
depend on
$\beta, \Oo$,
the only--if--part of the statement follows. $\bBox$

As has been pointed out in \cite[Sec.\ 5]{BuDaLo}, the quantities
$|| \TbO ||^p_p $
are a certain substitute for the partition function of the Gibbs
canonical ensemble at temperature
$( p \beta )^{-1}$
in a container of size proportional to
$\Oo$.
Thus the nature of the scaling limit is intimately
related to thermal properties of the underlying theory. It would
be desirable to clarify this relation further.

We conclude this section with a result pertaining to
the nuclearity properties
of the maps $\TNbO$.

\begin{Theorem} Consider a theory where the maps $\TbO$ are $p$--nuclear
for some $0 < p < 1/3$ and $\limsup_{\la \searrow 0} \, || \TlblO ||_p
< \infty$. Then the maps
$\TNbO$,
defined in relation (4.1), are $q$--nuclear for
$q > 2p / ( 2 - 3p )$,
and there holds
$$ || \TNbO ||_q \leq c \, \limsup_{\la \searrow 0} || \TlblO ||_p, $$
where $c$ depends only on $p, q$. \end{Theorem}
\noindent {\em Proof:\/} Given
$0 < p < 1/3$,
we pick $q, r$ such that
$1 > q > 2r/(2-r) > 2p/(2-3p)$.
Then
$r > p/(1-p)$,
so it follows from relation (4.7) and the first part of Lemma 2.1 that
$$ \NNbO \leq \limsup_{\la \searrow 0} \, \exp{ ( c \, || \TlblO ||^r_p
/ \, \varepsilon^r )}, \quad \varepsilon > 0,$$
and consequently
$$ \sup_{\varepsilon > 0} \, \varepsilon \, ( \ln{\NNbO} )^{1/r}
\leq c^{1/r} \, \limsup_{\la \searrow 0} \, || \TlblO ||_p. $$
Since
$r < 2/3$,
the statement then follows
from the last part of Lemma 2.1. $\bBox$

The preceding proposition provides the basis for a more detailed
investigation of the properties of the scaling limit theories.
It can be used, for example, to establish the so--called
{\em (distal) split--property\/} \cite{DoLo} in the scaling limit,
provided the underlying theory has decent phase--space properties. The
crucial
step is the demonstration that the maps
$\TNbO$
have certain specific nuclearity properties which can be expressed
in various ways \cite{BuWi, BuDaLo, BuDaFr, BuYn}. In view of the
preceding proposition these properties follow from corresponding
properties of the maps
$\TbO$
in the underlying theory. We refrain from stating here the
pertinent conditions and refer to the quoted publications.
\setcounter{section}{4}
\section{Concluding remarks}
\setcounter{equation}{0}
\noindent
Making use of the novel concept of scaling algebra, introduced in
\cite{BuVe}, we have established an interesting relation between
the phase space properties of a theory and the nature of its
scaling limit. It turned out that some rough information on the
number of degrees of freedom affiliated with certain specific
regions of phase space is sufficient to distinguish between
theories with a classical and (pure) quantum scaling limit.
Moreover, one can deduce rather precise information on the phase
space properties of the scaling limit from corresponding properties
of the underlying theory.

The present results are a promising step towards the general
understanding and the classification of the short distance
properties of local nets of observables. But there are still
many open problems. It is, for example, an intriguing question
under which circumstances a theory has a {\em unique\/} quantum
scaling limit, cf.\ Sec.\ 3. Phase space properties of the theory
seem to matter also in this context, but there do not yet exist
any definitive results. The uniqueness of the scaling limit has
been established so far only in certain models and in dilation
invariant theories \cite{BuVe}.

It would be instructive to have a supply of examples illustrating
also the various other possibilities appearing in the general
classification of the structure of scaling limits, such as theories
with a classical or degenerate scaling limit. A simple example
of the classical type ought to be the net which is obtained from a
generalized free field
$\phi ( x )$
with continuous mass spectrum according to the following
procedure: the local algebras corresponding to regions of diameter
$\la$
are generated by the fields
$\bBox^{n_\la}  \phi ( x )$,
where the numbers
$n_\la$
tend to infinity as
$\la$
tends to $0$. In this way one obtains a local, Poincar\'e--covariant
and weakly additive net. But, as is apparent from the
construction, the algebras corresponding to shrinking regions
contain, apart from multiples of the identity, operators with
rapidly worsening ultraviolet properties. As has been pointed out
in \cite[Sec.\ 4]{BuVe} one may expect that such nets have a
classical scaling limit. The model resembles to a certain extent
the situation in field theories without ultraviolet fixed point,
where one cannot remove the cutoff and proceed to point--like
fields.

A model with a degenerate scaling limit ought to
be the infinite tensor product theory constructed from free
scalar fields with masses
$m \in 2^{\mbox{\footnotesize ${\Bbb \, Z}$}} \, m_0$,
where
$m_0 > 0$ is fixed.
This theory is invariant under the subgroup
$2^{\mbox{\footnotesize ${\Bbb \, Z}$}}$
of dilations and consequently should coincide with one of its
scaling limits. But the set of scaling limit states of a
theory is invariant under arbitrary scaling transformations,
hence there should appear in the scaling limit also the theories with
scaled mass spectrum
$\mu \, 2^{\mbox{\footnotesize ${\Bbb \, Z}$}} \, m_0$,
which are in general non--isomorphic for different
$\mu > 0$. The details of
these simple but instructive examples will be worked out elsewhere.

It would also be desirable to understand better the relation between
the present algebraic approach to the renormalization group
\cite{BuVe} and the conventional field theoretic treatment \cite{Zi}.
By a combination of these methods one may hope to gain new insights
into the possible ultraviolet properties of local field theories.
To this end it would be necessary to identify in the algebraic
setting invariants, in analogy to the beta function and the
anomalous dimensions of local operators
in field theory, which are apt to
describe in a more quantitative manner the ultraviolet
properties of a local net. A certain step in this direction is
the purely algebraic characterization of asymptotically free
theories, proposed in \cite[Sec.\ 4]{BuVe}.

Another interesting problem is the analysis of the
superselection and particle structure
emerging in the scaling limit. As
has been pointed out in \cite{Bu}, cf. also \cite{BuVe}, one can
identify the {\em ultraparticles\/}, i.e., the particle-like structures
appearing at small scales, such as quarks, gluons and leptons,
with the particle content (in the sense of Wigner) of the
scaling limit theory. Similarly, the {\em ultrasymmetries\/} of
a theory, i.e., the symmetries which are visible at small scales,
such as colour and flavour, can be identified with the global
gauge group of the scaling limit theory. The discussion
of interesting issues, such as the reconstruction of
the local gauge group from the local observables or the confinement
problem can then be based on a comparison of the particle and
symmetry content of a theory with the ultraparticle and
ultrasymmetry content of its scaling limit. Thus the concept
of scaling algebra seems also a useful tool for the
investigation of these more conceptual problems.
\\[24pt]
{\Large {\bf Acknowledgement}}
\\[18pt]
\noindent
The author gratefully acknowledges financial support
by the DFG (Deutsche Forschungsgemeinschaft).
\end{document}